\documentclass[twocolumn,showpacs,preprintnumbers,amsmath,amssymb,aps,prl]{revtex4}

\usepackage{graphicx}
\usepackage{dcolumn}
\usepackage{bm}

\begin{document}

\preprint{APS/123-QED}

\title{Damped Population Oscillation in a Spontaneously Decaying\\ Two-Level Atom Coupled to a Monochromatic Field}

\author{Sun Kyung Lee}
\author{Hai-Woong Lee}%
\affiliation{%
Department of Physics, Korea Advanced Institute of Science and Technology, Daejeon 305-701, Korea}

\date{\today}

\begin{abstract}
We investigate the time evolution of atomic population in a two-level atom driven by a monochromatic radiation field, taking spontaneous emission into account. The Rabi oscillation exhibits amplitude damping in time caused by spontaneous emission. We show that the semiclassical master equation leads in general to an overestimation of the damping rate and that a correct quantitative description of the damped Rabi oscillation can thus be obtained only with a full quantum mechanical theory. 
\end{abstract}

\pacs{pacs}
\maketitle

The atom-field interaction is one of the most fundamental problems of quantum optics \cite{zubairy,loudon,allen,scully}. Despite the success of the semiclassical approach which treats the atom quantum mechanically but the field classically, there exist many optical phenomena, an accurate description of which requires a full quantum-mechanical treatment of both the atom and the field. One such example is the collapse and revival \cite{eberly,rempe,brune,cirac,wineland,haroche} of the Rabi oscillation in the Jaynes-Cummings model \cite{jaynes}. 

The influence of dissipation (spontaneous decay, cavity damping ) on the collapse and revival of the Rabi oscillation has been studied in the past \cite{barnett, agarwal}. When one considers a transition in the microwave region where the first observation of the collapse and revival was made \cite{rempe}, spontaneous decay can certainly be neglected. The past investigations on the influence of the cavity damping on the collapse and revival have shown  that the collapse is not affected much by the cavity damping, but the revivals, especially at large times, suffer attenuation in the oscillation amplitude \cite{barnett, agarwal}. In the optical domain, spontaneous emission may be expected to play a much stronger role. It has been shown, however, that revivals in the optical region are still much more sensitive to cavity decay than to spontaneous emission \cite{quang}.   

In this letter we report on our study of the influence of spontaneous emission on the collapse and revival of the Rabi oscillation. The reason for our study of spontaneous emission, despite its relative unimportance compared with cavity decay, is that there exists a subtle quantum effect when the collapse and revival phenomenon is combined with spontaneous decay. The quantum effect has to do with the damping rate of the amplitude of the Rabi oscillation. As is well known, the amplitude of the Rabi oscillation decreases progressively at each successive revival, even in the absence of spontaneous decay and any other dissipation, because a smaller and smaller number of dephased oscillations rephase at each successive revival time. Spontaneous decay works to further decrease the oscillation amplitude over the already existing decrease in its absence. We are in this work mainly concerned with the damping rate of the oscillation amplitude due to spontaneous emission. We show that the semiclassical theory tends to overestimate this damping rate. A correct quantitative description of the damped Rabi oscillation is thus obtained in general only with a full quantum mechanical treatment. The overestimation stems from the fact that the semiclassical picture of the transitions involved is oversimplified, as will be seen below.

\begin{figure}[b]
\includegraphics[width=4cm]{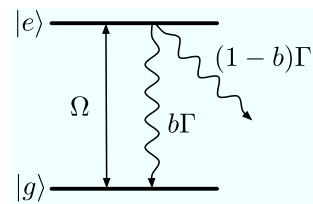}
\caption{\label{fig:one} Stimulated and spontaneous transitions of a two-level atom in the semiclassical picture. $\Omega$ is the Rabi frequency and $b\Gamma$ and $(1-b)\Gamma$ represent the spontaneous emission rate into the ground state and other external states, respectively.}
\end{figure}
Let us first look at the semiclassical treatment of a two-level atom driven by a resonant, monochromatic radiation field. The driving field induces the Rabi oscillation between the excited state $|e\rangle$  and the ground state $|g\rangle$ of the atom. We also take into consideration spontaneous decay of the excited state $|e\rangle$ into the ground state $|g\rangle$ and other external states. The transitions that the atom undergoes are schematically shown in Fig.~{\ref{fig:one}}. The master equation for the density operator $\rho$ of the atom is given in the interaction picture under the dipole and rotating wave approximations by 
\begin{eqnarray}
\frac{\partial\rho}{\partial t} = &&-i[\,\Omega\sigma_{+}+\Omega^{\ast}\sigma_{-},\rho \,] \nonumber \\
&& +\frac{\Gamma}{2}(2\,b\,{\sigma}_{-}\rho\,{\sigma}_{+}-\sigma_{+}\sigma_{-}\rho - \rho\,\sigma_{+}\sigma_{-}),
\label{eq:one}
\end{eqnarray}
where $\Omega$ is the Rabi frequency, $\sigma_{+}$ and $\sigma_{-}$ are the atomic raising and lowering operators given respectively by $\sigma_{+}= |e \rangle \langle g|$ and $ \,\sigma_{-}=|g \rangle \langle e|$, $\Gamma$ represents the total spontaneous decay rate of the excited state, and the parameter $b$ ($ 0 \leq b \leq 1$) is the branching ratio into the ground state, i.e., the decay rate into the ground state is $b\Gamma$ and that into all other external state is $(1-b)\Gamma$. 

From Eq.~(\ref{eq:one}) we immediately obtain the optical Bloch equations for the density matrix elements $\rho_{gg}=\langle g |\rho| g\rangle$, $\rho_{ee}=\langle e |\rho| e\rangle$,  and $\rho_{ge}=\langle g|\rho| e\rangle$, 
\begin{subequations}
\label{eq:two}
\begin{eqnarray}
&\dot{\rho_{gg}}= &-i\,(\Omega ^{\ast}\rho _{eg}-\Omega\,\rho _{ge}) + b\Gamma\rho_{ee},\\
&\dot{\rho_{ee}}= &-i\,(\Omega\,\rho _{ge}-\Omega ^{\ast}\rho _{eg}) - \Gamma\rho_{ee},\\
&\dot{\rho_{ge}}= & -i\, \Omega^{\ast}(\rho_{ee}-\rho_{gg})-\frac{\Gamma}{2}\rho_{ge}.
\end{eqnarray}
\end{subequations}
Eqs.~(\ref{eq:two}) yield a third-order differential equation for $\rho_{ee}$ which can be solved analytically to give 
\begin{eqnarray}
&\rho_{ee}(t) =& e^{-\frac{\Gamma}{2}t}e^{f_{1}t}\frac{8|\Omega |^{2}} {3({3f_{1}}^2+{f_2}^{2})}[(1 - \,e^{\frac{-3f_{1}}{2}t}\nonumber \\
&&\times(\sqrt{3}\,\frac{f_{1}}{f_{2}}\sin{\frac{\sqrt{3}}{2}f_{2}t}+\cos{\frac{\sqrt{3}}{2}f_{2}t})],
\label{eq:three}
\end{eqnarray} 
where 
\begin{subequations}
\label{eq:four}
\begin{eqnarray}
&f_{1}\,=& \frac{A^{1/3}}{6}-\frac{16|\Omega |^{2}-\Gamma^{2}}{2A^{1/3}},\\
&f_{2}\,=& \frac{A^{1/3}}{6}+\frac{16|\Omega |^{2}-\Gamma^{2}}{2A^{1/3}},\\
&A \,= & 216\,b\Gamma\,{|\Omega |}^{2}+3\,\sqrt {5184\,{b}^{2}{\Gamma}^{2}{|\Omega |}^{4}+3
\left( 16\,|\Omega |^{2}-{\Gamma}^{2}\right)^{3}}.\nonumber\\
\end{eqnarray}
\end{subequations}
In obtaining Eq.~(\ref{eq:three}), we have assumed that the atom is initially in its ground state; $\rho_{gg}(0)=1$ and $\rho_{ee}(0)=\rho_{ge}(0)=0$. 

From now on we limit ourselves to the strong coupling regime in which $\Omega$ is sufficiently greater than $\Gamma$ and, as a result, both $f_1$ and $f_2$ are real and positive. Eq.~(\ref{eq:three}) indicates that, in the strong coupling regime, the excited state population exhibits a damped oscillatory behavior with a frequency of $\frac{\sqrt{3}}{2}f_2$ and a damping rate of $(\Gamma +f_1)/2$. Our calculation indicates that in general $f_1$ increases roughly linearly with $b$ while $f_2$ does not depend much on $b$, which means that the damping rate of the population oscillation increases with $b$ while the oscillation frequency is roughly independent of $b$. 

As a special case we set $b=0$ in Eq.~(\ref{eq:three}) and obtain 
\begin{equation}
\rho _{ee}(t)= \frac{4|\Omega |^{2}}{{\zeta}^2} e^{-\frac{\Gamma}{2}t}\sin ^{2}{\frac{\zeta}{2}t},
\label{eq:five}
\end{equation}
where $\zeta = \sqrt{4 |\Omega |^{2}-\frac{\Gamma ^{2}}{4}}$. In the other extreme case in which $b=1$, we obtain 
\begin{equation}
\rho _{ee}(t)= \frac{4|\Omega | ^{2}}{8|\Omega |^{2}+\Gamma ^{2}}[1-(\cos{\lambda t}+\frac{3\Gamma}{4\lambda}\sin{\lambda t})\,e^{-\frac{3}{4}\Gamma t}], 
\label{eq:six}
\end{equation}
where, $\lambda = \sqrt{4|\Omega |^{2}-\frac{\Gamma ^2}{16}}$. In the limit $|\Omega| \gg \Gamma$, Eq.~(\ref{eq:six}) becomes 
\begin{equation}
\rho _{ee}(t) \cong \frac{1}{2}(1-\cos{(2|\Omega |t)}e^{-\frac{3}{4}\Gamma t}). 
\label{eq:seven}
\end{equation}
The solutions, Eqs.~(\ref{eq:five}) and (\ref{eq:six}), have been given previously \cite{loudon,scully} although, to our knowledge, the general solution, Eq.~(\ref{eq:three}), has not. One sees from Eqs.(\ref{eq:five})-(\ref{eq:seven}) that the semiclassical treatment predicts that the damping rate of the population oscillation increases with respect to the branching ratio $b$ from $\frac{\Gamma}{2}$ at $b=0$ to $\frac{3}{4}\Gamma$ at $b=1$.   

Why does the population oscillation damp out faster when the branching ratio $b$ is greater, i.e., when the spontaneous decay to the ground state has a higher weight? The clue to this question can be provided by examining the role played by the last term $b\Gamma\rho_{ee}$ of Eq.~(\ref{eq:two}a). Roughly speaking, this term represents a kind of pumping of population into the ground state. This ``internal pumping" into the ground state, being proportional to $\rho_{ee}$, is strongest when the excited state population is maximum, i.e., when the ground state population is minimum. The transition from the excited state to the ground state occurs because the excited state population is greater, but this ``internal pumping" due to the term $b\Gamma\rho_{ee}$ works in the direction to spoil the very reason for the transition. As a result, the greater this term $b\Gamma\rho_{ee}$ is, the faster the population oscillation dies out. 

We now present a full quantum-mechanical description treating the driving field as well as the atom quantum mechanically. The master equation for the density operator $\rho$ of the atom-field system reads 
\begin{eqnarray}
\frac{\partial\rho}{\partial t} = &&-i[\,g\sigma_{+}a+g^{\ast}\sigma_{-}a^{\dagger},\rho\,] \nonumber \\
&& +\frac{\Gamma}{2}(2\,{b}\,{\sigma}_{-}\rho\,{\sigma}_{+}-\sigma_{+}\sigma_{-}\rho - \rho\,\sigma_{+}\sigma_{-}),
\label{eq:eight}
\end{eqnarray}
where $a ^{\dagger}$ and $a$ are, respectively, the creation and annihilation operators of a photon of the driving field mode, and $g$ is the parameter that measures the atom-field coupling (not to be confused with the ground state). 

The optical Bloch equations now take the form 
\begin{subequations}
\begin{eqnarray}
\dot{\rho}_{{gn+1,gn+1}}&=&  -i\sqrt{n+1}\,(g^{\ast}\rho _{en,gn+1}-g\,\rho _{gn+1,en})\nonumber \\ &&+ b\Gamma\rho_{en+1\,en+1},\\
\dot{\rho}_{{en,en}} &=& -i\sqrt{n+1}\,(g\,\rho _{gn+1,en}-g^{\ast}\rho _{en,gn+1}) \nonumber \\ && - \Gamma\rho_{en,en},\\
\dot{\rho}_{{gn+1,en}}&=& -i\sqrt{n+1}\, g^{\ast}(\rho_{en,en}-\rho_{gn+1,gn+1}) \nonumber \\ && -\frac{\Gamma}{2}\rho_{gn+1,en},
\end{eqnarray}
\label{eq:nine}
\end{subequations}
where $\rho_{gn+1,gn+1} =\langle g,n+1|\,\rho\, |g,n+1\rangle$, $\rho_{en,en}=\langle e,n|\, \rho\, |e,n\rangle$, $\rho_{gn+1,en}=\langle g,n+1|\, \rho\, |e,n\rangle$, and $|e,n\rangle$ represents the state of the atom-field system in which the atom is in the excited state and the field has $n$ photons, and similarly for $|g,n+1\rangle$. 

\begin{figure}
\includegraphics[width=6cm]{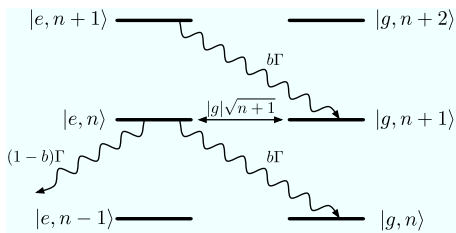}
\caption{\label{fig:three} Stimulated and spontaneous transitions of the atom-field system in the quantum picture. Only the transitions that directly influence the populations of the states $|e,n\rangle$ and $|g,n+1\rangle$ are shown. $|g|$ is the parameter that measures the atom-field coupling, and $b\Gamma$ and $(1-b)\Gamma$ represent the spontaneous emission rate into the ground state and other external states, respectively. }
\end{figure}
The transitions that are represented by Eqs.~(\ref{eq:nine}) are depicted in Fig.~\ref{fig:three}. We emphasize that the transition between $|g,n+1\rangle$ and $|e,n\rangle$ is not closed even for $b=1$, because spontaneous emission of a photon from $|e,n+1\rangle$ increases the population of the state $|g,n+1\rangle$, and similarly spontaneous emission of a photon into $|g,n\rangle$ as well as into external states decreases the population of the state $|e,n\rangle$. This ``openness" of the quantum transitions is responsible for the quantum effect we discuss in the following.

Eqs.~(\ref{eq:nine}) cannot be solved analytically in general, because, as mentioned above, the transition between $|g,n+1\rangle$ and $|e,n\rangle$ is not completely closed. Only in the limit $b=0$, Eqs.~(\ref{eq:nine}) are reduced to a closed set of equations relating $\rho_{gn+1,\,gn+1}$, $\rho_{en,\,en}$, and $\rho_{gn+1,\,en}$, and an analytic solution for $\rho _{ee}(t)=\sum_{n=0}^{\infty}\rho_{en,en}(t)$ can be obtained which reads 
\begin{equation}
\rho _{ee}(t)=\sum_{n=0}^{\infty} \frac{4|g|^{2}(n+1)e^{-|\alpha |^2}\alpha ^{2(n+1)}}{(n+1)!|\xi_{n} |^2}
e^{-\frac{\Gamma}{2}t}\sin^{2}{\frac{\xi_{n}}{2}t}
\label{eq:ten}
\end{equation}
where $\xi_{n} = \sqrt{4|g|^{2}(n+1)-\frac{\Gamma ^2}{4}}$. In obtaining Eq.~(\ref{eq:ten}), we have assumed that initially the atom is in the ground state and the driving field in the coherent state $|\alpha\rangle$, i.e., $\rho_{gn+1,\,gn+1}(0)=\frac{e^{-|\alpha |^2}\alpha ^{2(n+1)}}{(n+1)!}$, $\rho_{en,\,en}(0)=\rho_{gn+1,\,en}(0)=0$. Comparing Eq.~(\ref{eq:ten}) with its semiclassical counterpart, Eq.~(\ref{eq:five}), we first note that Eq.~(\ref{eq:ten}) consists of the summation of an infinitely large number of terms arising from the fact that the Rabi frequency for the transition between $|g,n+1\rangle$ and $|e,n\rangle$ is dependent on $n$. As is well known, this leads to the collapse and revival of the Rabi oscillation \cite{eberly, rempe, brune, cirac, wineland, haroche} . Our main concern in this work is however not the collapse and revival but the damping rate of the population oscillation. Eq.~(\ref{eq:ten}) indicates that this damping rate for the case $b=0$ is $\frac{\Gamma}{2}$, in agreement with the semiclassical damping rate of Eq.~(\ref{eq:five}). The question then arises: does the damping rate calculated according to the quantum theory agree with the semiclassical damping rate also for nonzero values of $b$?, i.e., does the quantum damping rate increase with the branching ratio $b$ as predicted by the semiclassical theory? This is the main issue we wish to address ourselves in this work. 

We show in the following that the answer to the above question is no and that the quantum damping rate of the population oscillation does not vary much with the branching ratio $b$. In order to see this, let us first recall that the reason for the increase of the damping rate with respect to the branching ratio in the semiclassical theory is that the ``internal pumping" represented by the last term $b\Gamma\rho_{ee}$ of Eq.~(\ref{eq:two}a) works to weaken the oscillation. Similarly, we see from Eq.~(\ref{eq:nine}a) that the last term $b\Gamma\rho_{en+1,\,en+1}$ plays the role of pumping population into the state $|g,n+1\rangle$. The strength of this ``pumping" is determined by the population of the state $|e,n+1\rangle$, while the state $|g,n+1\rangle$ into which population is pumped undergoes the Rabi oscillation with the state $|e,n\rangle$ not with the state $|e,n+1\rangle$. Since there is no definite phase relationship between the population of the state $|e,n+1\rangle$ and that of the state $|e,n\rangle$(or $|g,n+1\rangle$), this pumping strengthens the oscillation as much as it weakens the oscillation. After a sufficiently long time, therefore, the effect of the pumping term can largely be neglected. We thus expect that the damping rate of the population oscillation at any nonzero value of $b$ is roughly the same as that at $b=0$, i.e., it is $\frac{\Gamma}{2}$ regardless of $b$. This expectation based on the quantum-mechanical consideration of transitions contrasts sharply with the semiclassical prediction. 
\begin{figure}
\begin{tabular}{cc}
\includegraphics[width=0.22\textwidth]{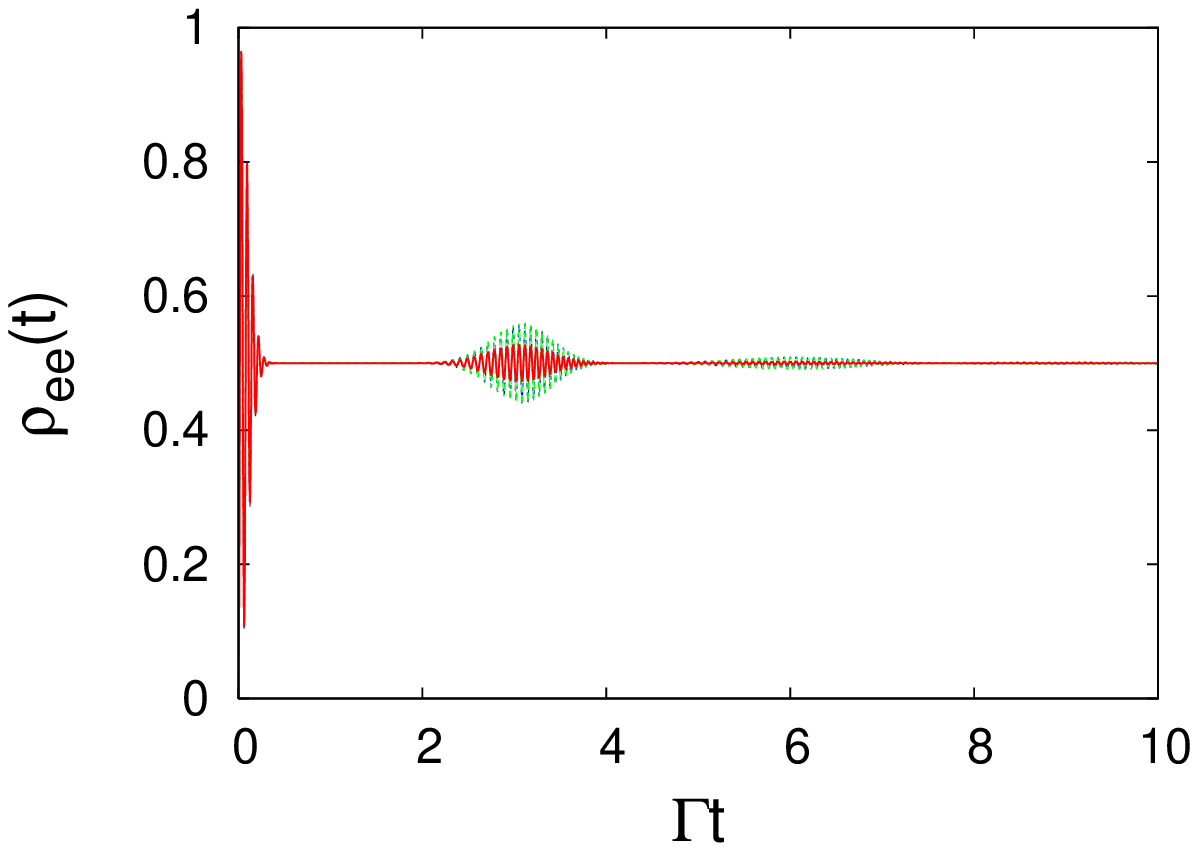} & 
\includegraphics[width=0.22\textwidth]{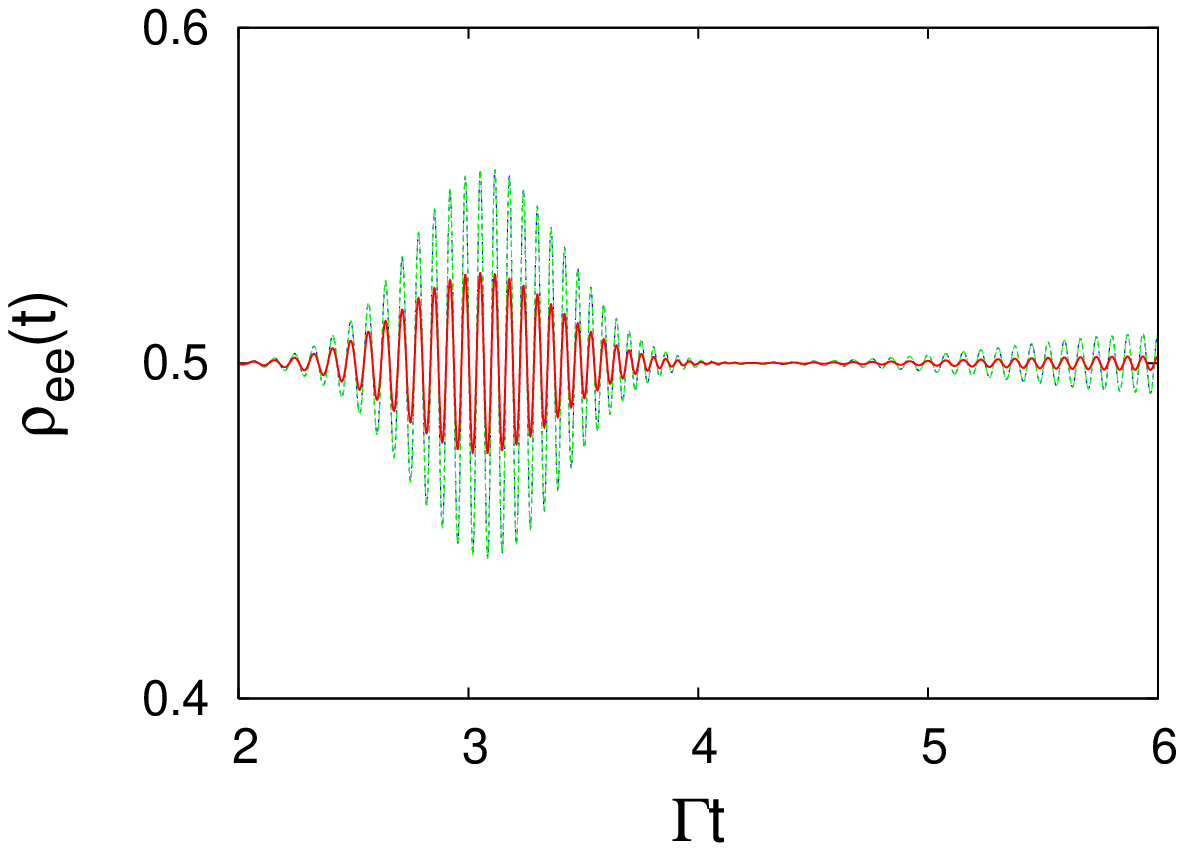} \\
(a) & (b) \\
\end{tabular}
\caption{\label{fig:four} Variation of the excited state population with time. (a) $\rho_{ee}^{exact}(t)$(green-dotted), $\rho_{ee}^{semi}(t)$(red-solid), and $\rho_{ee}^{quant}(t)$(blue-dashed) as a function of time for the case $|g|=10\Gamma$, $\langle n \rangle = |\alpha|^2 = 25$ and $b=1$. $\rho_{ee}^{exact}(t)$ and $\rho_{ee}^{quant}(t)$ are indistinguishable. (b) an expanded view of the revival region.}
\end{figure}

We now present numerical data to support the above expectation. The ``exact" solution for $\rho_{ee}(t)$, which we denote by $\rho_{ee}^{exact}(t)$, was obtained by numerically integrating the optical Bloch equations, Eqs.~(\ref{eq:nine}). The system parameters we chose for our computation are $|g|=10\Gamma$, $\langle n \rangle = |\alpha|^2 =25$, and $b=1$. The ratio $\frac{|g|}{\Gamma}=10$ can be achieved in an optical cavity \cite{kimble, spillane}. The value $b=1$ is chosen because at this value the semiclassical damping rate takes on a maximum value of $\frac{3}{4}\Gamma$ and differs most from the expected quantum damping rate of $\frac{\Gamma}{2}$. In Fig.~\ref{fig:four}(a) we plot $\rho_{ee}^{exact}$ as a fuction of time. We try to fit this exact curve with an approximate formula using a guessed value $\Gamma_{guess}$ of the damping rate. We take the approximate formula as 
\begin{eqnarray}
\rho_{ee}^{approx} (t) & \cong & \frac{1}{2} \biggl( 1-\sum_{n=0}^{\infty}\frac{e^{-|\alpha |^2}\alpha ^{2(n+1)}\cos{(2|g|\sqrt{n+1}t)} } {2(n+1)!} \nonumber \\
&&\times e^{-\Gamma_{guess}t } \biggr). 
\label{eq:eleven}
\end{eqnarray}
Eq.~(\ref{eq:eleven}) is suggested as a quantum counterpart of the semiclassical solution, Eq.~(\ref{eq:seven}), on the basis of comparison of the semiclassical and quantum solutions for the case $b=0$, Eq.~(\ref{eq:five}) and Eq.~(\ref{eq:ten}). With $\Gamma_{guess}$ accurately guessed, Eq.~(\ref{eq:eleven}) is expected to closely approximate $\rho_{ee}^{exact}(t)$, as long as $t$ is not extremely large. At very large times, Eq.~(\ref{eq:eleven}) predicts that $\rho_{ee}^{approx}(t \rightarrow \infty)\rightarrow\frac{1}{2}$, whereas the exact solution should yield $\rho_{ee}^{exact} (t\rightarrow\infty)\rightarrow 0$ because of the irreversible ``leak" into the lowest energy state $|g,0\rangle$ (see Fig.~\ref{fig:three}). A similar type of leak was discussed in an investigation of the collapse and revival in the absence of dissipation \cite{knight}. 

\begin{figure}[t]
\includegraphics[width=4cm]{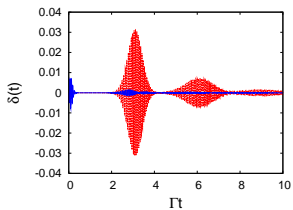}
\caption{\label{fig:five} The differences $\delta^{semi}(t)=\rho_{ee}^{exact}(t)-\rho_{ee}^{semi}(t)$(red-dotted) and $\delta^{quant}(t)=\rho_{ee}^{exact}(t)-\rho_{ee}^{quant}(t)$(blue-solid) for the case $|g|=10\Gamma$, $\langle n \rangle = |\alpha|^2= 25$ and $b=1$. }
\end{figure}
The best guessed value for $\Gamma_{guess}$ on the basis of the quantum argument given above is $\Gamma_{guess}=\frac{1}{2}\Gamma$. On the other hand, the semiclassical theory predicts $\Gamma_{guess}=\frac{3}{4}\Gamma$. The approximate solutions $\rho_{ee}^{quant}(t)$ and $\rho_{ee}^{semi}(t)$, which are obtained by substituting $\Gamma_{guess}=\frac{1}{2}{\Gamma}$ and $\Gamma_{guess}=\frac{3}{4}\Gamma$, respectively, into Eq.~(\ref{eq:eleven}), are plotted in Fig.~\ref{fig:four}(a) along with the exact numerical solution $\rho_{ee}^{exact} (t)$. We note that, on the scale with which this figure is drawn, $\rho_{ee}^{quant}(t)$ is indistinguishable from $\rho_{ee}^{exact}(t)$, whereas $\rho_{ee}^{semi}(t)$ clearly exhibits a faster damping than $\rho_{ee}^{exact}(t)$ and $\rho_{ee}^{quan}(t)$. Since the effect of different damping rates shows up most clearly in the revival region, we show in Fig.~\ref{fig:four}(b) an expanded view of the three curves, $\rho_{ee}^{exact}(t)$,  $\rho_{ee}^{semi}(t)$ and $\rho_{ee}^{quant}(t)$, in the first and second revival regions.  Even in this expanded figure, the two curves $\rho_{ee}^{exact}(t)$ and $\rho_{ee}^{quant}(t)$ are indistinguishable.  

As a further comparison we plot in Fig.~\ref{fig:five} the differences $\delta^{semi}(t) = \rho_{ee}^{exact}(t)-\rho_{ee}^{semi}(t)$ and $\delta^{quant}(t) = \rho_{ee}^{exact}(t)-\rho_{ee}^{quant}(t)$. While $\delta^{semi}(t)$ reaches a value as large as $3\times 10^{-2}$,  $\delta^{quant}(t)$ remains within $10^{-3}$. We conclude therefore that the correct damping rate for the case $b=1$ is $\frac{\Gamma}{2}$, in disagreement with the semiclassical prediction of $\frac{3}{4}\Gamma$. 

In conclusion we have shown that the Rabi oscillation exhibited by a two-level atom interacting with a monochromatic field damps out with the rate of $\frac{\Gamma}{2}$, regardless of the branching ratio $b$ of the spontaneous decay into the ground state, where $\Gamma$ represents the total decay rate of the excited state. This is in contradiction to the semiclassical prediction that the damping rate increases with $b$ from $\frac{\Gamma}{2}$ at $b=0$ to $\frac{3}{4}\Gamma$ at $b=1$. The reason for the constant damping rate lies in the openness of the quantum transition between the two states $|g,n+1\rangle$ and $|e,n\rangle$. The oversimplified structure of the transitions employed in the semiclassical theory, as depicted in Fig.~\ref{fig:one}, cannot correctly describe this openness which is clearly visible in the full quantum mechanical structure shown in Fig.~\ref{fig:three}. The effect of a smaller quantum damping rate compared with the corresponding semiclassical damping rate shows up most clearly in the amplitude of the revived Rabi oscillations.

This research was supported by the ``Single Quantum-Based Metrology in Nanoscale" project of the Korea Research Institute of Standards and Science(KRISS). 



\end{document}